# Determination of the superhump period of the dwarf nova V701 Tauri during the 2005 December superoutburst

Jeremy Shears & David Boyd

We report new measurements of the superhump period of the UGSU-type dwarf nova V701 Tau during the 2005 December superoutburst. Using unfiltered time series CCD observations on 3 nights, we determine a probable superhump period $P_{sh}$ = 0.0690 ± 0.0002d, but note that our data also permit a possible shorter period of 0.0663 ± 0.0002d. The longer period agrees with the value measured during the first recorded superoutburst in 1995.

## Background

Dwarf novae are a class of cataclysmic variables, which are known to be interacting binary stars where a cool main sequence star (the secondary) loses mass to a white dwarf primary.[1] The result is the formation of an accretion disc around the primary. From time to time the accretion disc flips from a cooler, dimmer state, to a hotter, brighter state, resulting in what we see as an outburst, in which the star brightens by several magnitudes. Dwarf novae of the SU UMa family (UGSU) occasionally exhibit *superoutbursts*, which are typically 0.5 to 1 magnitude brighter than normal outbursts and last up to 10 times longer. During a superoutburst the light curve of a UGSU star is characterised by *superhumps*. These are modulations which are a few percent longer than the orbital period and are thought to be caused by precession of the accretion disc.[2]

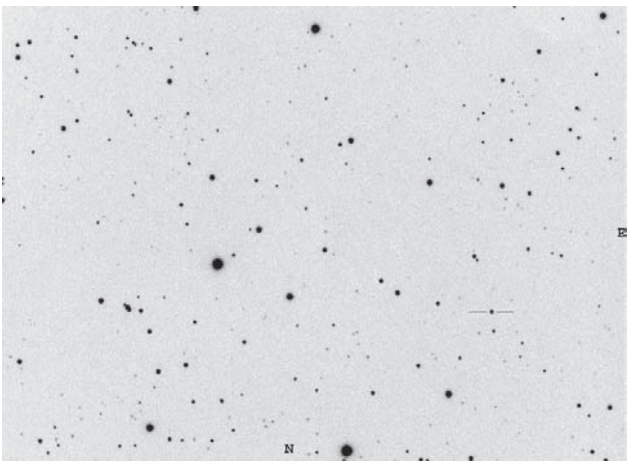

**Figure 1.** V701 Tau in outburst at 15.1C. 2005 Dec 6, 22:11UT. Takahashi FS102, 0.1m refractor. 60sec image with unfiltered Starlight Xpress SXV-M7 CCD. Field 25'×19' with South at the top. *(Jeremy Shears)*

V701 Tau was discovered by Esastova on 1970 Sep 2 at a photographic magnitude of 15.00.[3] Esastova followed the outburst until Sep 12, by which time the star had faded to 15.97. The next attempted observation was on Sep 19, but the star was no longer detectable (<16.5).

The next recorded outburst of V701 Tau was detected visually on Christmas day 1995 by Tonny Vanmunster at 14.5v and confirmed by Gary Poyner.[4] Time resolved V-band photometry was conducted by Taichi Kato at the Ouda Station in Japan using the 0.6m telescope on several nights. Data obtained on Dec 27 revealed the presence of small amplitude (0.05 to 0.06 mag in V) superhumps,[5] establishing for the first time that V701 Tau is a dwarf nova belonging to the UGSU family. By Dec 29 the superhumps were fully developed, having a V amplitude of 0.25 mag. Superhumps were also detected on several subsequent nights, allowing a superhump period, $P_{sh}$, to be deduced of 0.0689 days.[5] The final positive observation was on 1996 Jan 13, at 19.5V, eighteen days after the outburst was reported.

The magnitude of V701 Tau at quiescence is not known, but the General Catalogue of Variable Stars states that it is below 21.[6] This suggests the amplitude range is at least 6.5 mag.

## V701 Tau and the Recurrent Objects Programme

V701 Tau was added to the Variable Star Section's Recurrent Objects Programme (ROP) in 2001.[7] The aim of the ROP is to encourage monitoring of poorly characterised dwarf novae (and other stars). Charts for V701 Tau are available from the AAVSO web site.[8] An outburst was reported by JS on 2005 Dec 6 at 15.1C (Figure 1) and confirmed visually by Gary Poyner at 14.9v, following a telephone alert. This appears to be the first outburst detected since that of 1995 and only the third outburst ever recorded.

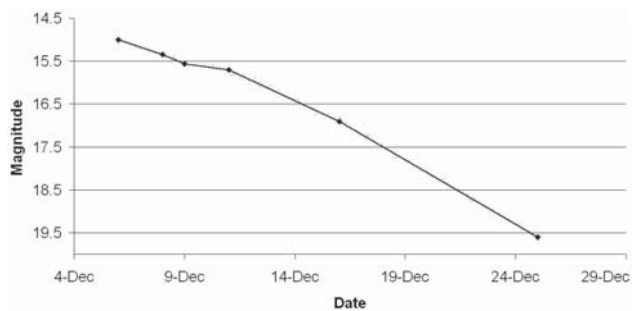

**Figure 2.** Light curve of the 2005 Dec outburst. Magnitudes are unfiltered CCD. *(Jeremy Shears and David Boyd)*

A light curve for the 2005 outburst is shown in Figure 2. The last positive observation was on Dec 25 at 19.6C, nineteen days after the outburst was detected. The decline was almost linear, with an average decay of 0.24 mag/day.

## Time resolved photometry

Time resolved photometry was conducted by JS on 2005 Dec 6, starting at 22:11 UT for a period of 3 hr. The instrument was a 0.1m Takahashi FS102 apochromatic refractor with an unfiltered Starlight Xpress SXV-M7 CCD camera. During this time 158 unfiltered images were collected, each having an integration time of 60s. The raw images were dark-subtracted and flat-fielded before being analysed photometrically using the 'multiple-images photometry' function of the *AIP4WIN* software.[9] The comparison star used in the photometry was '121' (magnitude 12.1) on the AAVSO chart. The resulting light curve is shown in Figure 3. Errors represent combined estimates of statistical uncertainty and instrumental noise. An inspection of the light curve reveals superhumps, with a peak-to-trough amplitude of approximately 0.3 mag. The fact that such well developed superhumps were evident suggests that the superoutburst was already well advanced when the outburst was discovered.

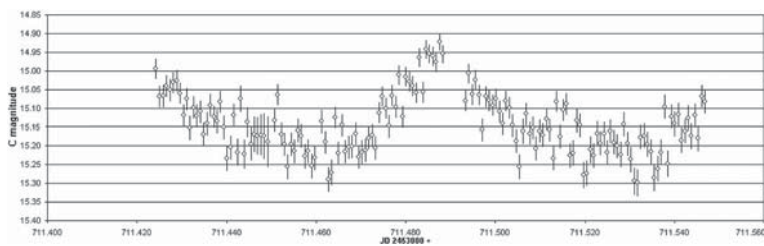

**Figure 3.** Superhumps in V701 Tau on 2005 Dec 6. *(Jeremy Shears)*

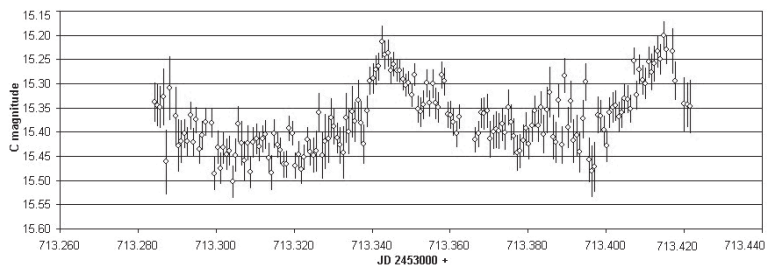

**Figure 4.** V701 Tau light curve on 2005 Dec 8. *(David Boyd)*

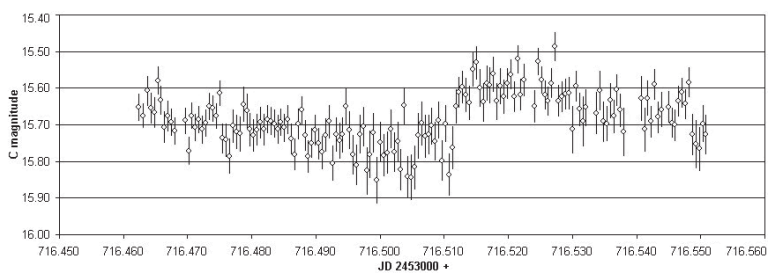

**Figure 5.** V701 Tau light curve on 2005 Dec 11. *(David Boyd)*

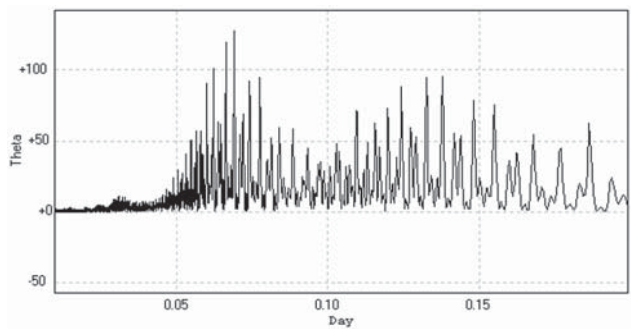

**Figure 6.** Power spectrum of V701 Tau during the 2005 Dec outburst.

Photometry was also carried out by DB on 2005 Dec 8 (3.3 hr), 9 (0.3 hr), 11 (2.1 hr) and 25 (0.75hr). The instrument was a 0.35m SCT with an unfiltered Starlight Xpress SXV-H9 CCD camera. Integration times on the first 3 nights were 40s or 60s depending on conditions. The measurement on the final night was based on a stacked integration of 40×60s. A similar procedure was used to analyse the images, in this case using comparison star 144. Light curves for Dec 8 and 11 are shown in Figures 4 and 5. They also show prominent superhumps, but with lower amplitude than on Dec 6.

## Period analysis

We performed period analysis on the integrated time series data from Dec 6, 8 and 11 using algorithms in the *Peranso* software package.[10] We find the highest peak in the power spectrum at a period of 0.0690 ± 0.0002d, which agrees with the value obtained during the 1995 outburst, but note the presence of a slightly lower peak at a period of 0.0663 ± 0.0002d. (Other peaks are aliases of one or other of these periods.) Both these periods show good phase coherence over our integrated dataset. The power spectrum using the ANOVA algorithm is shown in Figure 6 and the corresponding phase diagrams obtained by folding the data on the shorter and longer periods are given in Figures 7 and 8 respectively. We therefore conclude that our observations indicate a probable superhump period, $P_{sh} = 0.0690 ± 0.0002$d, but also permit the possibility of a shorter period of $0.0663 ± 0.0002$d. The quoted uncertain-ties are calculated using the Schwarzenberg–Czerny method.[11] Observations during future superoutbursts are needed to confirm the superhump period.

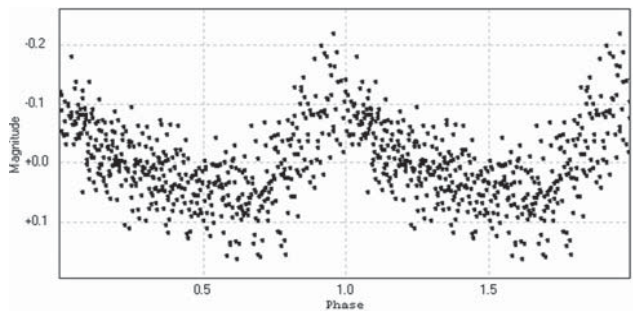

**Figure 7.** Phase diagram based on a period of 0.0663d.

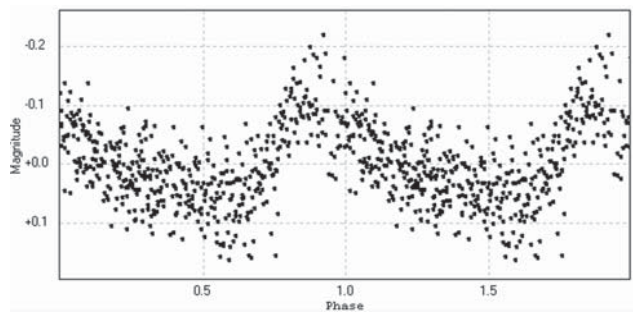

**Figure 8.** Phase diagram based on a period of 0.0690d.

ments during the preparation of this paper, especially for drawing to our attention the VSNET data on this star.

**Addresses: JS:** 'Pemberton', School Lane, Bunbury, Tarporley, Cheshire CW6 9NR [bunburyobservatory@hotmail.com]
**DB:** 5 Silver Lane, West Challow, Wantage, Oxon. OX12 9TX [drsboyd@dsl.pipex.com]

# Conclusion

During the 2005 December superoutburst of the UGSU-type dwarf nova V701 Tau, we carried out unfiltered time series CCD observations on 3 nights, which showed superhumps to be present in the light curve. Analysis of these data yield a probable superhump period $P_{sh} = 0.0690 \pm 0.0002$d but also permit a possible shorter period of $0.0663 \pm 0.0002$d. The longer period agrees with the value measured during the first recorded superoutburst in 1995.

# Acknowledgments

The authors would like to thank Gary Poyner, coordinator of the ROP, for his encouragement and for his helpful com-